# Graded index SCH transistor laser: Analysis of various confinement structures

Mohammad Hosseini[1*], Hassan Kaatuzian[1**], and Iman Taghavi[2***]

[1] Photonics Research Laboratory, Electrical Engineering Department, Amirkabir University of Technology, Hafez Ave., Tehran, Iran

[2] Electrical and Computer Engineering Department, Georgia Institute of Technology, Atlanta, GA, 30332

*smohammad.hsi@aut.ac.ir; **hsnkato@aut.ac.ir; ***staghavi3@gatech.edu

New configuration of confinement structure is utilized to improve optoelectronic performances, including threshold current, AC current gain as well as optical bandwidth and optical output power of single quantum well transistor laser. Considering the drift component in addition to the diffusion term in electron current density, a new continuity equation is developed to analyze the proposed structures. Physical parameters including, electron mobility, recombination lifetime, optical confinement factor, electron capture time and photon lifetime is calculated for new structures. Based on solving continuity equation in separate confinement heterostructures, threshold current reduces 67% and optical output power increases 37% when graded index layers of $Al_\xi Ga_{1-\xi}As$ ($\xi$: 0.05→0 in left side of quantum well, $\xi$: 0→0.02 in right side of quantum well) are used instead of uniform GaAs in base region.

Carrier transport across the separate confinement heterostructure (SCH) has substantial effects on dc and ac characteristics of quantum well (QW) lasers [1]. Also, carrier population of quantum well excited states, which depends on confinement structure, plays an important role in the threshold specifications of SCH and GRIN (graded-index) SCH quantum well lasers [2]. Heterojunction Bipolar Transistor Laser (HBTL) is a type of quantum well lasers, containing one (or a few) quantum well (s) and two SCHs in its base region, where carrier dynamics is governed by various structural factors. The TL that we study here is based on npn HBT (n-InGaP/p-GaAs/n-GaAs) [3]. In GaAs-based HBTs, the common approach to establish quasi electric field is linearly grading the aluminum (or indium) content of AlGaAs (or InGaAs) [4]. In the present letter, we employ graded index layers of $Al_\xi Ga_{1-\xi}As$ instead of simply uniform GaAs in the base region to achieve various level of confinements.

Energy band diagram of the primary TL proposed by Feng and Holonyak [3] under forward bias is depicted in Fig. 1 (a), named the first structure in this letter. In the first proposed design, called the second structure, graded index layers of $Al_\xi Ga_{1-\xi}As$ ($\xi$: 0.05 → 0 in SCH1, $\xi$: 0 → 0.02 in SCH2) are used, (Fig. 1 (b)). In the third structure, graded index layers of $Al_\xi Ga_{1-\xi}As$ ($\xi$: 0.07 → 0.02 in SCH1, $\xi$: 0.02 → 0 in SCH2) are used (Fig. 1 (c)). Calculating slope of the conduction band profile we gain an estimate for the induced built-in potential in SCH1, 2.

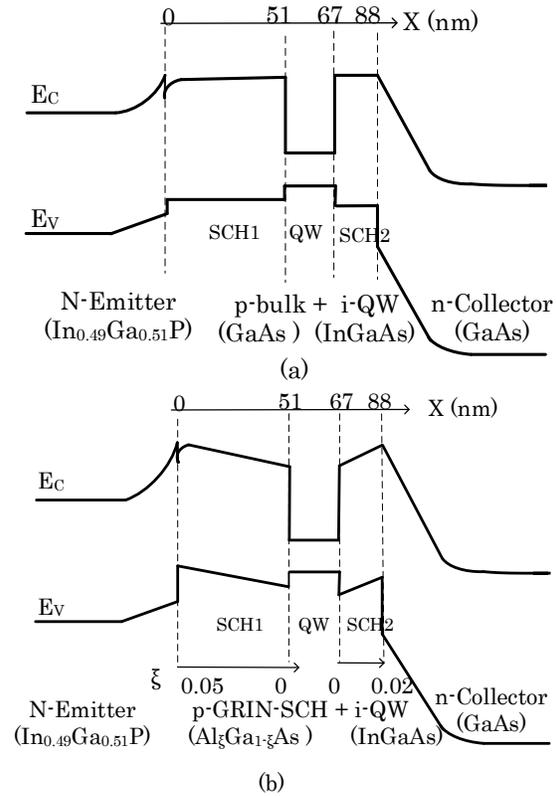

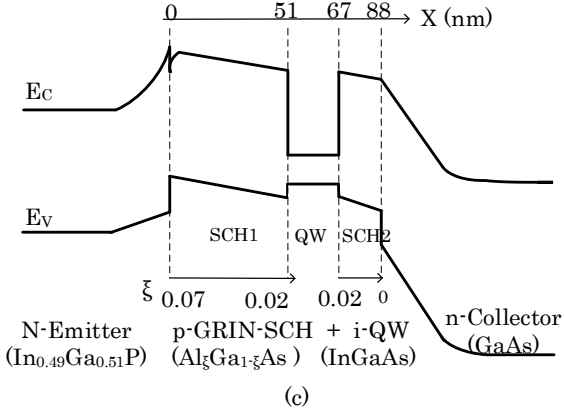

Fig. 1 Schematic energy band diagram of (a) primary TL proposed by Feng and Holonyak [3] (1st structure), (b) 1st proposed GRIN-SCH structure, (2nd structure) and (c) 2nd proposed GRIN-SCH structure (3rd structure), under forward bias. Slope of the band diagram in SCH1, 2 determines the magnitude of quasi electric field.

To analyze the proposed structures, equation for the electron current density, including both drift and diffusion components, can be written as [1]

$$J_n = qD_n \left( \frac{qn\varepsilon}{kT} + \frac{\partial n}{\partial x} \right) \quad (1)$$

where n is the base electron concentration, D is the diffusion constant, $\varepsilon = \Delta E_g/qW_{SCH}$ is the base quasi electric field (induced built in potential), $\Delta E_g$ denotes the bandgap energy difference at two sides of SCH regions, and $W_{SCH}$ is the SCH width. The continuity equation describing injected minority electron population transporting from the emitter to the collector is:

$$\frac{\partial n}{\partial t} = \frac{1}{q}\frac{\partial J_n}{\partial x} - \frac{n}{\tau_B} \quad (2)$$

where $\tau_B$ is the base recombination lifetime. Combining Eqs. (1) and (2), following equations for the SCH1 and SCH2 regions can be obtained under steady-state condition ($\partial n/\partial t = 0$)

$$D_1 \frac{\partial^2 n}{\partial x^2} + \mu_1 \varepsilon_1 \frac{\partial n}{\partial x} - \frac{n}{\tau_{B1}} = 0 \quad \text{SCH1} \quad (3)$$

$$D_2 \frac{\partial^2 n}{\partial x^2} + \mu_2 \varepsilon_2 \frac{\partial n}{\partial x} - \frac{n}{\tau_{B2}} = 0 \quad \text{SCH2} \quad (4)$$

where $\mu$ is the minority electron mobility. Boundary conditions needed to solve Eqs. (3) and (4) are $n(W_B) = 0$ and $n(W_{qw}^-) = n(W_{qw}^+) = N_{V.S}$, where $N_{V.S}$ is the carrier density of the virtual states (VS) located at $x = W_{qw}$. Current density of the virtual state ($J_{V.S}$) including both the drift and diffusion components is:

$$J_{V.S} = \underbrace{qD_1 \frac{\partial n}{\partial x}(W_{qw}^-) - qD_2 \frac{\partial n}{\partial x}(W_{qw}^+)}_{\text{diffusion component}} + \underbrace{q\mu_1 n(W_{qw}^-)\varepsilon_{B1} - q\mu_2 n(W_{qw}^+)\varepsilon_{B2}}_{\text{drift component}} \quad (5)$$

Equations (6) and (7) make appropriate connection between carrier densities and current densities related to the quantum well ($n_{QW}, j_{QW}$) and VS as [5]:

$$\frac{j_{V.S.}}{qd} = \frac{j_{QW}}{qd} - \frac{n_{V.S.}}{\tau_S} \quad (6)$$

$$\frac{j_{QW}}{qd} = \frac{n_{V.S.}}{\tau_{cap}} - \frac{n_{QW}}{\tau_{esc}} \quad (7)$$

In Eqs. (6) and (7), d is the QW width, $\tau_S = 1000\ ps$ is the spontaneous emission lifetime, $\tau_{esc} = 10\ ps$ is the escape lifetime from the QW to the VS. $\tau_{cap}$ is the overall capture lifetime for the QW [6].
Finally, emitter and collector current densities are found by

$$J_E = \underbrace{qD_1 \frac{\partial n}{\partial x}}_{\text{diffusion component}} + \underbrace{q\mu_1 n(x)\varepsilon_1}_{\text{drift component}} \quad ; \quad x = 0 \quad (8)$$

$$J_C = \underbrace{qD_2 \frac{\partial n}{\partial x}}_{\text{diffusion component}} + \underbrace{q\mu_2 n(x)\varepsilon_2}_{\text{drift component}} \quad ; \quad x = W_B \quad (9)$$

The semiconductor optical gain function is defined as [7]

$$G(N_{QW}, S_0) = \frac{G_0(N_{QW} - N_{tr})}{1 + \epsilon S_0} \quad (10)$$

where $N_{tr}$ is the transparency electron density, $S_0$ is the dc value of photon concentration, $\epsilon = 1.5 \times 10^{-17} cm^3$ is the gain compression factor, and $G_0 = 1 \times 10^{-5} cm^3 s^{-1}$ is the differential gain of the active layer. The threshold current is calculated using the optical confinement factor ($\Gamma$) and optical loss ($\alpha$) of each structure [2],

$$\Gamma G_{th} = \alpha + (1/2L)\ln(1/R_1 R_2) \quad (11)$$

where L = 450 $\mu m$ is the laser cavity length and $R_1$ = 0.32 and $R_2$ = 0.32 are the facet reflectivity. From the calculated value of $G_{th}$ we deduce the corresponding $N_{tr}$.

Minority electron mobility in highly doped $Al_\xi Ga_{1-\xi}As$ were calculated in [8], while other physical parameters of Multiple Quantum Well (MQW) TL, including diffusion constant, carrier lifetimes and optical confinement factor are calculated by Taghavi et al. [6]. All of physical parameters for the new proposed structures are calculated and summarized in Table 1.

Table 1. Calculated physical parameters for all three structures.

| Device parameter | | Symbol | Unit | Calculation Approach Ref. | Structure | | |
|---|---|---|---|---|---|---|---|
| | | | | | 1st | 2nd | 3rd |
| Effective minority electron mobility | SCH1 | $\mu_1$ | cm²/Vs | [8] | 1068 | 964.7 | 919.325 |
| | SCH2 | $\mu_2$ | cm²/Vs | | 1068 | 1014.05 | 1014.05 |
| Diffusion constant | SCH1 | $D_1$ | cm²/s | [4] | 27.61 | 24.88 | 23.71 |
| | SCH2 | $D_2$ | cm²/s | | 27.61 | 26.16 | 26.16 |
| Quasi electric field | SCH1 | $\varepsilon_1$ | V/cm | [4] | 0 | 1.22×10⁴ | 1.22×10⁴ |
| | SCH2 | $\varepsilon_2$ | V/cm | | 0 | 1.18×10⁴ | 1.18×10⁴ |
| Effective recombination lifetime | SCH1 | $\tau_{B1}$ | ps | [6] | 201 | 220 | 243 |
| | SCH2 | $\tau_{B2}$ | ps | | 201 | 210 | 210 |
| Optical confinement factor | | $\Gamma$ | % | [6] | 5.82 | 5.65 | 5.51 |
| Optical loss | | $\alpha_i$ | cm⁻¹ | [6] | 20.14 | 19.56 | 19.07 |
| Photon lifetime | | $\tau_p$ | ps | [6] | 2.57 | 2.59 | 2.61 |
| Electron capture time | | $\tau_{cap}$ | ps | [6] | 0.90 | 0.56 | 0.57 |

Solving Eqs. (3) and (4) for increasing base current, minority electron density in the base region can be calculated as shown in Fig. (2). Compared with calculated electron distribution for standard SQW TL in [3, 9], excited electron population in QW is larger in the first proposed design (Fig. 2 (a)). Fig. 2 (a) is in good agreement with the calculated carrier population in the barrier layer for the linear GRIN-SCH quantum well laser discussed in [10]. Fig. 2 (b) also comply with the results of graded-base HBTs discussed in [4].

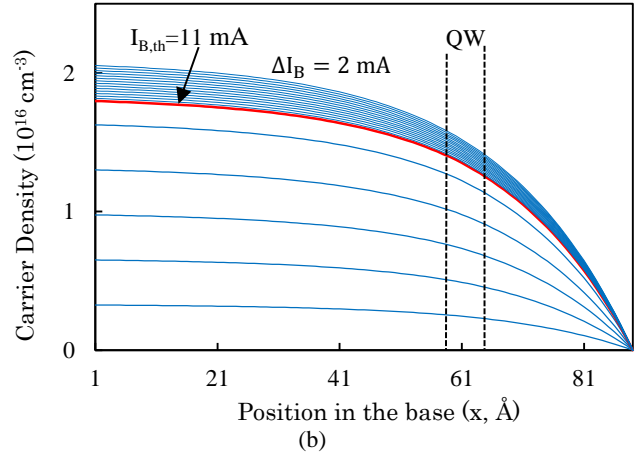

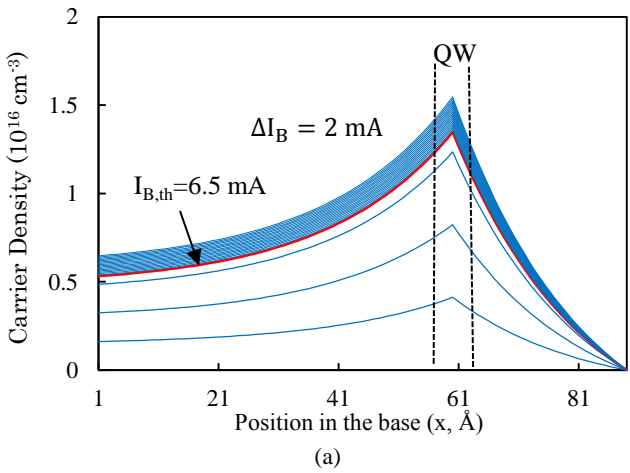

Fig. 2 Calculated minority electron distribution.

Fig. 3 shows the variation of dc current gain ($\beta_{dc} = I_C/I_B$) for all three structures. DC current gain reduces significantly, when second structure is employed. As shown in Fig. 2, the number of captured electrons in QW is larger compared with other structures, as a result, the proportion of electrons captured by collector is smaller, therefore we should expect a reduction in the current gain.

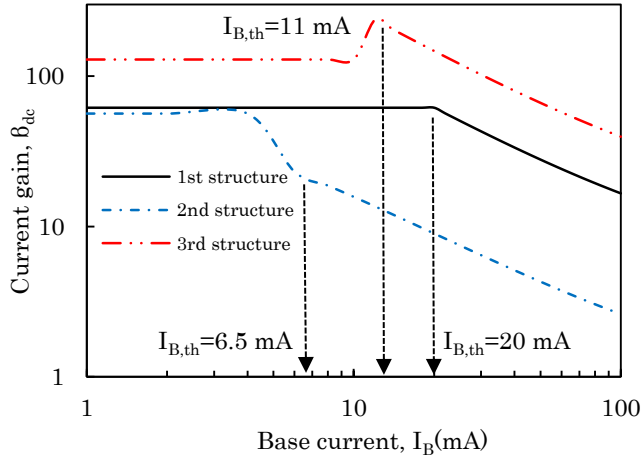

Fig. 3 Calculated dc current gain, $\beta_{dc}(I_C/I_B)$.

Optical output power for different base currents are calculated for SQW TL in [11]. Calculated values of light output power for GRIN-SCH and standard SQW TL are shown in Fig. 4. As can be seen, the optical output power for the same base current is larger for GRIN-SCH TL compared with standard SQW TL.

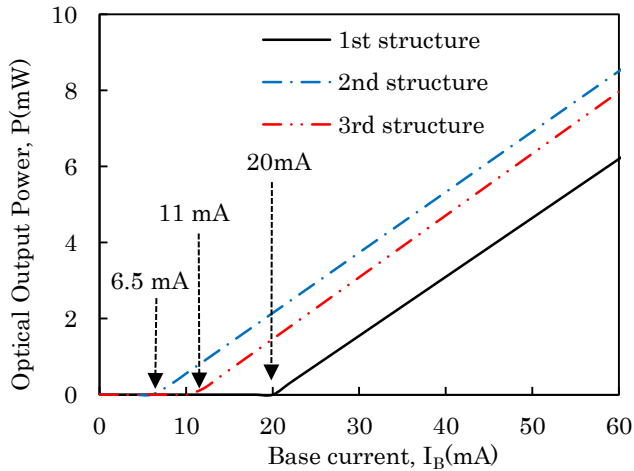

Fig. 4 Calculated Optical output power, P.

Fig. 5 displays the optical frequency response of SQW TL for different configurations (i.e. structures). One can result when graded index layers of $Al_\xi Ga_{1-\xi}As$ is utilized, electron transport time from emitter to QW reduces because of tilt in the band diagram (Fig. (1) (b) and (c)). Since that electron transit time is critical to frequency response operation, so optical bandwidth increases for GRIN-SCH structures, compared with traditional, i.e. non-graded, structures.

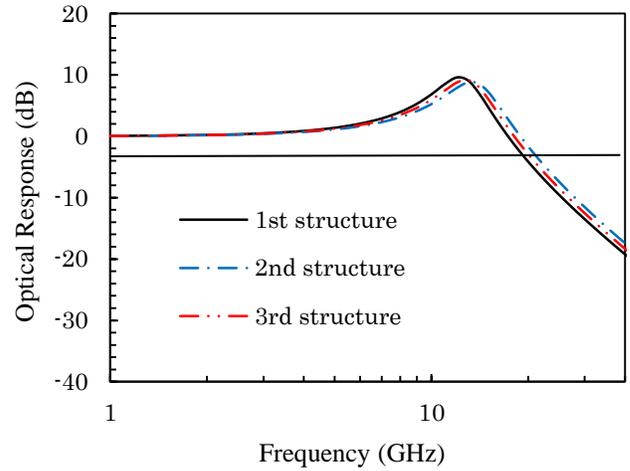

Fig. 5 Optical frequency response for different structures.

Fig. 6 exhibit optoelectronic performances, including threshold current, AC current gain, optical bandwidth and optical power output. For optimum results in an integrated optical link or microprocessor, we need switching devices with threshold currents as low as possible while optical bandwidth is as high as possible. According to our simulations, structure 2 could be of more interest. Other characteristics, e.g. current gain and power output, can also play important roles for specific purposes which requires other tailored structures.

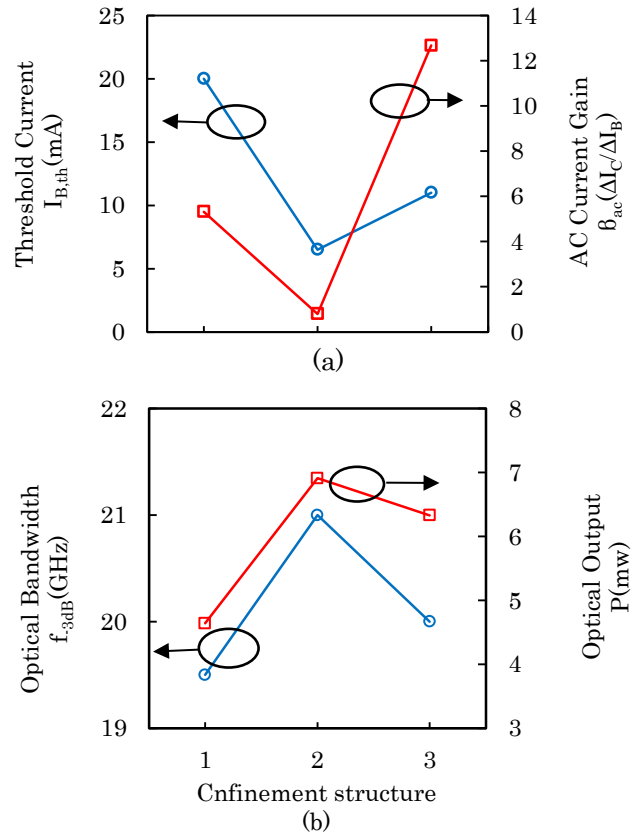

Fig. 6 Confinement structure dependency, (a) Electrical. (b) Optical characteristics of the SQW TL.

In this letter, we theoretically studied the effect of replacing simply uniform GaAs base region with graded index layers of $Al_\xi Ga_{1-\xi}As$. Considering the effect of internal field due to this graded base, we developed a new continuity equation and carefully calculated all required physical parameters. Eventually, we have shown that if $Al_\xi Ga_{1-\xi}As$ ($\xi$: 0.05→0 in SCH1, $\xi$: 0→0.02 in SCH2) is utilized as the base region, the threshold current reduces 67% and optical output power increases 37% compared with the previous structure with uniform base region. Additionally, the optical bandwidth is anticipated to improve up to 21 GHz.